\documentclass[pra,twocolumn,showpacs]{revtex4}
\usepackage{amsfonts}
\usepackage{amssymb}
\usepackage{amsmath}
\usepackage{graphicx}

\begin{document}

\title{Generation of frequency multiplexed entangled single photons assisted
by the entanglement}
\author{Chun-Hua Yuan$^{1}$, L. Q. Chen$^{1}$, Z. Y. Ou$^{1,2,*}$, Weiping
Zhang$^{1,\dag}$}
\affiliation{$^{1}$State Key Laboratory of Precision Spectroscopy, Department of Physics,
East China Normal University, Shanghai 200062, P. R. China\\
$^{2}$Department of Physics, Indiana University-Purdue University
Indianapolis, 402 North Blackford Street, Indianapolis, Indiana 46202, USA}
\date{\today }

\begin{abstract}
We present a scheme to generate the frequency multiplexed entangled (FME)
single photons based on the entanglement between two species atomic mixture
ensemble. The write and reads fields driven according to a certain timing sequence,
the generation of FME single photons can be repeated until success is
achieved. The source might have significant applications in wavelength
division multiplexing quantum key distribution.
\end{abstract}

\maketitle

In classical communication, wavelength division multiplexing (WDM) is a
successful method for increasing the transmission capacity of optical fiber
systems \cite{Ishio,Keiser}. In quantum communication, photon is the basic
information carrier. To increase the amount of information, we may use the
similar technique, but different channels may not be independent to each
other \cite{Ramelow}. A source for the wavelength division multiplexing
quantum key distribution (WDM-QKD) system must satisfy two conditions: (1)
it must be a true single-photon source (to avoid eavesdropping); (2) it must
contain photons with different wavelengths \cite{Yabushita,Fan}. A frequency
multiplexed entangled (FME) single photon state has the form of (two
frequency components)
\begin{equation}
|1_{\omega _{1},\omega _{2}}\rangle =c_{1}|1\rangle _{\omega _{1}}|0\rangle
_{\omega _{2}}+c_{2}|0\rangle _{\omega _{1}}|1\rangle _{\omega _{2}},
\end{equation}%
which can be prepared in our scheme and might be an excellent candidate for
WDM-QKD.

The generation of two- and multi-color light fields has been widely studied
in three-level \cite{Wanare,Li2006,Zhang2007,Yang2010} and double (or
multi-) $\Lambda $ configuration atomic systems \cite%
{Cerboneschi,Raczy,Li,Chong,Kang,Moiseev}. In these systems rich spectral
features have been manifested, due to constructive or destructive quantum
interference between different channels. For three-level atomic system, the
studies of multi-color field generation have been reported theoretically
\cite{Wanare,Li2006} and experimentally \cite{Zhang2007,Yang2010}. Extending
the analysis to a double (or multi-) $\Lambda $ system \cite%
{Cerboneschi,Raczy,Li,Chong,Kang,Moiseev}, it is possible to simultaneously
generate two- and multi-color light fields in the medium based on a dark
state polariton (DSP) consisting of low-lying atomic excitations and photon
states of two frequencies. The existence of the DSP in the $\Lambda $\ and
double (or multi-) $\Lambda $ atomic system required that the fields obey
certain conditions for frequency, amplitude, and phase matchings. If the
phase of the two (or multi-) color reading light fields is mismatched, then
one of the pulses will be absorbed and lost \cite{Zhang2007,Yang2010,Kang}.
Here, we present a scheme that the generation of frequency entangled single
photons is from two species atomic mixture ensemble without phase-matching
requirement, and it is also a natural extension of existing work. Another
potential approach to prepare the FME photon state based on coherent
conversion was presented very recently \cite{Chen}.

\begin{figure}[tbp]
\centerline{\includegraphics[scale=0.55,angle=0]{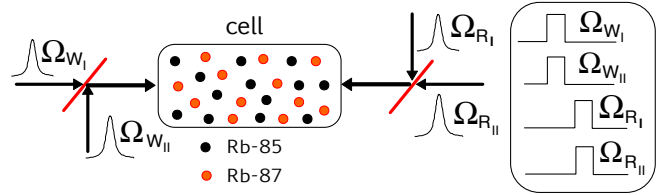}}
\caption{(Color online) An isotope mixture ensemble in the cell. $\Omega
_{W_{\text{I}}}$, $\Omega _{W_{\text{II}}}$: the Rabi frequencies of the
write fields; $\Omega _{R_{\text{I}}}$, $\Omega _{R_{\text{II}}}$: the Rabi
frequencies of the read fields. Inset: timing sequence.}
\label{fig1}
\end{figure}

In this brief report, we propose a scheme to generate the FME single photons
based on the entanglement between two species atomic mixture ensemble. Here
we first use the single-photon interference to realize the entanglement
between two species atomic mixture ensemble, as the method is used to
realize the entanglement between two identically distant atomic ensemble
\cite{Duan}, but it is extension to two different species atoms. Based on
the generated entanglement, then we can deterministically obtain the FME
single photon .

\begin{figure}[tbp]
\centerline{\includegraphics[scale=0.5,angle=0]{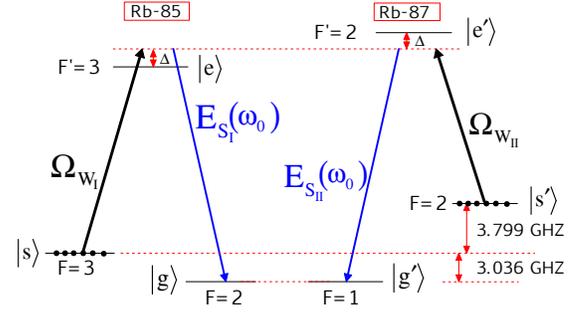}}
\caption{(Color online) Energy level diagrams of two species atoms $^{85}$Rb
and $^{87}$Rb, respectively. The Stokes fields $E_{S_{\text{I}}}(\protect%
\omega _{0})$ and $E_{S_{\text{II}}}(\protect\omega _{0})$ are generated by
the write fields $\Omega _{W_{\text{I}}}$ and $\Omega _{W_{\text{II}}}$,
respectively. $\Delta $ is the detuning.}
\label{fig2}
\end{figure}

\begin{figure}[tbp]
\centerline{\includegraphics[scale=0.5,angle=0]{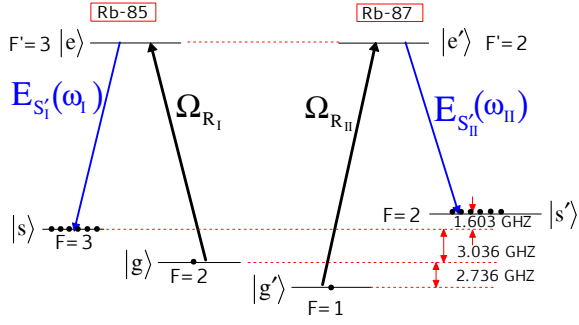}}
\caption{(Color online) The Stokes fields $E_{S_{\text{I}}^{\prime}}(\protect%
\omega _{\text{I}})$ and $E_{S_{\text{II}}^{\prime}}(\protect\omega _{\text{%
II}})$ are generated by the read fields $\Omega _{R_{\text{I}}}$ and $\Omega
_{R_{\text{II}}}$, respectively. }
\label{fig3}
\end{figure}

\begin{figure}[tbp]
\centerline{\includegraphics[scale=0.5,angle=0]{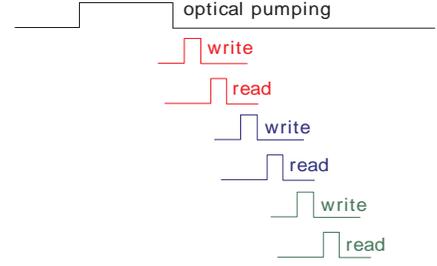}}
\caption{(Color online) The cycled pulse sequences. }
\label{fig4}
\end{figure}

The outline of our scheme is as follows. We consider an isotope mixture
medium, containing $N_{\text{I \ }}$and $N_{\text{II}}$ (I, II denote two
different isotope) $\Lambda $-type two species atoms shown in Fig.~\ref{fig1}%
. Here we consider a $^{85}$Rb+$^{87}$Rb mixed atomic ensemble \cite{Note},
and Fig.~\ref{fig2} shows their energy level diagrams. To generate the FME
single photons, the system firstly is driven by the pumping fields and takes
place the Raman scattering processes, then is driven by two read fields.
There are four steps for this process. Step $1$: optical pumping prepares
all mixed atoms in the respective ground hyperfine levels $|s\rangle $ and $%
|s^{\prime }\rangle $, and the whole state of the atomic ensemble and the
Stokes fields can now be written as
\begin{equation}
|\psi (0)\rangle =|c^{(0)}\rangle _{\text{I}}|c^{(0)}\rangle _{\text{II}%
}|0\rangle _{S_{\text{I}}}|0\rangle _{S_{\text{II}}},  \label{eq1}
\end{equation}%
where $|c^{(0)}\rangle _{\text{I}}\equiv |s_{1},...,s_{N_{\text{I}}}\rangle $
and $|c^{(0)}\rangle _{\text{II}}\equiv |s_{1}^{\prime },...,s_{N_{\text{II}%
}}^{\prime }\rangle $ are collective atomic ground states, and $|0\rangle
_{S_{\text{I}}}$ and $|0\rangle _{S_{\text{II}}}$ are the vacuum states of
the Stokes fields $E_{S_{\text{I}}}$\ and $E_{S_{\text{II}}}$, respectively.
Step $2$: sending in two off-resonant write fields $\Omega _{W_{\text{I}}}$
and $\Omega _{W_{\text{II}}}$ interacting with the transition $|s\rangle
\rightarrow |e\rangle $ and the transition $|s^{\prime }\rangle \rightarrow
|e^{\prime }\rangle $, respectively, and the Stokes fields $E_{S_{\text{I}}}$%
\ and $E_{S_{\text{II}}}$ with respective frequencies $\omega _{S_{\text{I}%
}}=\omega _{W_{\text{I}}}+\omega _{sg}$ and $\omega _{S_{\text{II}}}=\omega
_{W_{\text{II}}}+\omega _{s^{\prime }g^{\prime }}$\ are produced by
spontaneous Raman process from each of the isotopes as illustrated in Fig.~%
\ref{fig2}. We assume that the light-atom interaction time $\tau $ is short
or the intensities of the write fields are weak, so both the mean photon
numbers in the two forward-scattered Stokes pulse $E_{S_{\text{I}}}$\ and $%
E_{S_{\text{II}}}$ are much smaller than $1$. To erase the
distinguishability of the photons from the Stokes fields $E_{S_{\text{I}}}$\
and $E_{S_{\text{II}}}$, we choose the condition that the Stokes fields $%
E_{S_{\text{I}}}$\ and $E_{S_{\text{II}}}$ have the same frequencies, i.e., $%
\omega _{S_{\text{I}}}=\omega _{S_{\text{II}}}$ $(=\omega _{0})$. Step $3$:
when only one photon is detected by the detector, a single spin wave
excitation in each one of the two atomic species in the same region of space
is generated. Because the frequencies of the two Stokes fields from the
isotope are the same, we can not decide which isotope emit the photon, and
the final state of the mixed ensemble is a superposition of two
probabilities, i.e. an entangled state. Note that the generations of the
entangled state from our scheme and from the DLCZ scheme \cite{Duan} are
based on the single-photon interference. But the mechanism are completely
different. The former is based on the inner interference due to two Stokes
fields use a common vacuum state. The latter is external interference based
on a beam splitter where two Stokes fields are generated from different
vacuum state. Step $4$: after generation of the entangled state of the mixed
ensemble, another two read fields $\Omega _{R_{\text{1}}}$ and $\Omega _{R_{%
\text{II}}}$ are driven on the isotope mixture system, and the spin wave is
coherently converted into the photon with frequency $\omega _{\text{I}}$ or $%
\omega _{\text{II}}$ shown in Fig.~\ref{fig3}. After driven by two read
fields $\Omega _{R_{\text{1}}}$ and $\Omega _{R_{\text{II}}}$, the
two-atomic-isotope ensemble radiates photons and both transitions lead to
the same final atomic state of the whole ensemble $|c^{(0)}\rangle _{\text{I}%
}|c^{(0)}\rangle _{\text{II}}$, then the FME single photon will appear. The
generation efficiency could be improved with the use of a recycled pulsed
sequence shown in Fig. \ref{fig4}, which will be described in detailed later.

In order to illustrate the main idea, we give a detailed analysis of the
generation process. Consider two species atomic mixture ensemble as shown in
Fig.~\ref{fig1}: the write fields $\Omega _{W_{\text{I}}}$ and $\Omega _{W_{%
\text{II}}}$ inject the atomic ensemble, and the quantized Stokes fields are
generated (see Fig.~\ref{fig2}). Because the single photons $|1_{\omega
_{0}}\rangle _{S_{\text{I}}}$ and $|1_{\omega _{0}}\rangle _{S_{\text{II}}}$
are indistinguishable, i.e., $|1_{\omega _{0}}\rangle _{S_{\text{I}%
}}=|1_{\omega _{0}}\rangle _{S_{\text{II}}}\equiv |1_{\omega _{0}}\rangle
_{S}$, then in this case the Stokes fields can be described by a common
operator. For conceptual simplicity we assume the quantized field
corresponds to a single mode of a running-wave cavity that is initially in a
vacuum state. Note that this cavity field consideration is valid in the
limit of unity finesse, i.e., in free space configuration. Then the
Hamiltonian for two species atomic mixture ensemble in the rotating frame
can be split into two parts $\hat{H}=\hat{H}_{\text{I}}+\hat{H}_{\text{II}}$
according to different isotopes
\begin{eqnarray}
\hat{H}_{\text{I}} &=&-\hbar \Delta \hat{\Sigma}_{ee}+[\hbar \Omega _{W_{%
\text{I}}}\hat{\Sigma}_{es}+\hbar g_{\text{I}}\hat{a}\hat{\Sigma}_{{eg}}+%
\text{H.c.}], \\
\hat{H}_{\text{II}} &=&\hbar \Delta \hat{\Sigma}_{e^{\prime }e^{\prime
}}+[\hbar \Omega _{W_{\text{II}}}\hat{\Sigma}_{{e}^{\prime }{s}^{\prime
}}+\hbar g_{\text{II}}\hat{a}\hat{\Sigma}_{{e}^{\prime }{g}^{\prime }}+\text{%
H.c.}],
\end{eqnarray}%
where $\hat{\Sigma}_{\mu \nu }=\sum_{j}|\mu \rangle \langle \nu |$ are
collective atomic operators corresponding to transitions between atomic
states $|\mu \rangle $, $|\nu \rangle $, and $\Delta $ is the detuning. $g_{%
\text{I}}$ and $g_{\text{II}}$ are the atom-cavity field coupling constants
of two isotope, respectively. $\Omega _{W_{\text{I}}}$ and $\Omega _{W_{%
\text{II}}}$ are the respective Rabi frequencies of two write fields. The
evolution of atomic operators is then described by Heisenberg-Langevin
equations
\begin{equation}
\dot{\Sigma}_{\mu \nu }=-\gamma _{\mu \nu }\hat{\Sigma}_{\mu \nu }+\frac{i}{%
\hbar }[H,\hat{\Sigma}_{\mu \nu }]+\hat{F}_{\mu \nu },
\end{equation}%
where $\gamma _{\mu \nu }$ is a decay rate of coherence $|\mu \rangle
\longrightarrow |\nu \rangle $, and $\hat{F}_{\mu \nu }$ are associated
noise operators which have zero average and are correlated with $\delta $\
associated diffusion coefficients. We can make the adiabatic approximation
according to large single-photon detuning $\Delta \gg \gamma _{1}$ (let $%
\gamma _{eg}=\gamma _{es}=\gamma _{1}$), $\gamma _{2}$ (let $\gamma
_{e^{\prime }g^{\prime }}=\gamma _{e^{\prime }s^{\prime }}=\gamma _{2}$),
and first order in $\hat{a}$ ($\Sigma _{ss}\sim N_{\text{I}}$ and $\Sigma
_{s^{\prime }s^{\prime }}\sim N_{\text{II}}$), then we obtain the equations
of motion for the cavity mode and the ground state coherences $\hat{S}_{%
\text{I}}$ ($=\Sigma _{sg}/\sqrt{N_{\text{I}}}$) and $\hat{S}_{\text{II}}$ ($%
=\Sigma _{s^{\prime }g^{\prime }}/\sqrt{N_{\text{II}}}$)
\begin{eqnarray}
\dot{a} &=&-\kappa \hat{a}-i\chi _{\text{I}}\hat{S}_{\text{I}}^{\dag }-i\chi
_{\text{II}}\hat{S}_{\text{II}}^{\dag }+\hat{F}_{a}, \\
\dot{S}_{\text{I}}^{\dag } &=&-(\gamma _{gs}+\gamma _{L_{\text{I}}}+i\delta
_{L_{j}})\hat{S}_{\text{I}}^{\dag }+i\chi _{\text{I}}^{\ast }\hat{a}+\hat{F}%
_{S_{\text{I}}}, \\
\dot{S}_{\text{II}}^{\dag } &=&-(\gamma _{g^{\prime }s^{\prime }}+\gamma
_{L_{\text{II}}}-i\delta _{L_{j}})\hat{S}_{\text{II}}^{\dag }+i\chi _{\text{%
II}}^{\ast }\hat{a}+\hat{F}_{S_{\text{II}}},
\end{eqnarray}%
where $\kappa $ is the decay rate of the cavity mode, and $\chi _{j}=g_{j}%
\sqrt{N_{j}}\Omega _{W_{j}}/\Delta $ ($j=$I, II) is the coupling rate
between the collective spin excitation $\hat{S}_{j}$\ and the quantized
field $\hat{a}$, and$\ \gamma _{_{L_{j}}}=\gamma _{j}\left\vert \Omega
_{W_{j}}\right\vert ^{2}/\Delta ^{2}$ ($j=$I, II) is an optical pumping
rate, and $\delta _{L_{j}}=\left\vert \Omega _{W_{j}}\right\vert ^{2}/\Delta
$ ($j=$I, II) is the ac Stark shift.

In this case the evolution of the entire system is described by an effective
Hamiltonian
\begin{equation}
\hat{H}_{eff}=\hbar (\chi _{\text{I}}\hat{S}_{\text{I}}^{\dag }-\chi _{\text{%
II}}\hat{S}_{\text{II}}^{\dag })\hat{a}^{\dag }+\text{H.c..}
\end{equation}%
The time evolution of an initial state $|\psi (0)\rangle $ is described by
\begin{equation}
|\psi (t)\rangle =e^{-i[(\chi _{\text{I}}t\hat{S}_{\text{I}}^{\dag }-\chi _{%
\text{II}}t\hat{S}_{\text{II}}^{\dag })\hat{a}^{\dag }+\text{H.c.}]}|\psi
(0)\rangle .
\end{equation}%
After a short interaction time $\tau $, the coefficient $P_{j}=\chi _{j}\tau
$\ ($P_{j}\ll 1$, $j=$I, II) is very weak. The final output state of the
atomic collective mode and the forward-scattering Stokes mode can be
approximated as
\begin{eqnarray}
|\psi (\tau )\rangle  &=&|c^{(0)}\rangle _{\text{I}}|c^{(0)}\rangle _{\text{%
II}}|0\rangle -i[P_{\text{I}}|c^{(1)}\rangle _{\text{I}}|c^{(0)}\rangle _{%
\text{II}}  \notag \\
&&-P_{\text{II}}|c^{(0)}\rangle _{\text{I}}|c^{(1)}\rangle _{\text{II}%
}]|1_{\omega _{0}}\rangle _{S},
\end{eqnarray}%
where%
\begin{eqnarray}
|c^{(1)}\rangle _{\text{I}} &=&\frac{1}{\sqrt{N_{\text{I}}}}\sum_{l=1}^{N_{%
\text{I}}}e^{i(k_{\text{w}}^{(\text{I})}-\omega
_{0}/c)z_{j}}|s_{1},...,g_{l},...,s_{N_{\text{I}}}\rangle ,  \notag \\
|c^{(1)}\rangle _{\text{II}} &=&\frac{1}{\sqrt{N_{\text{II}}}}\sum_{l=1}^{N_{%
\text{II}}}e^{i(k_{\text{w}}^{(\text{II})}-\omega
_{0}/c)z_{j}}|s_{1}^{\prime },...,g_{l}^{\prime },...,s_{N_{\text{II}%
}}^{\prime }\rangle ,  \notag \\
&&
\end{eqnarray}%
and $k_{\text{w}}^{(j)}$ ($j=$I, II) is the wave vector of the write fields.
Here the terms with high order excitation are ignored, and the probability
amplitudes $P_{\text{I}}$ and $P_{\text{II}}$ are dependent on the atom
numbers $N_{\text{I \ }}$and $N_{\text{II}}$, respectively. A click on the
single-photon detector will project the atomic ensemble in the entangled
state
\begin{equation}
|\phi \rangle =\frac{1}{\sqrt{P_{\text{I}}^{2}+P_{\text{II}}^{2}}}\left[ P_{%
\text{I}}|c^{(1)}\rangle _{\text{I}}|c^{(0)}\rangle _{\text{II}}-P_{\text{II}%
}|c^{(0)}\rangle _{\text{I}}|c^{(1)}\rangle _{\text{II}}\right] .  \label{e8}
\end{equation}%
If $P_{\text{I}}=P_{\text{II}}$, the state is maximally entangled. After
generation of the entangled state of the two species atomic ensemble, we can
realize the FME single photons based on the generated entanglement of Eq. (%
\ref{e8}).

After generation of the entangled state of two species atomic ensemble, two
read fields $\Omega _{R_{\text{1}}}$ and $\Omega _{R_{\text{II}}}$\ nearly
resonant with respective transition $|g\rangle \rightarrow |e\rangle $ and $%
|g^{\prime }\rangle \rightarrow |e^{\prime }\rangle $ are driven on the
atomic ensemble, and the spin wave is coherently converted into the
anti-Stokes photon with frequency $\omega _{\text{I}}$ or $\omega _{\text{II}%
}$ (see Fig.~\ref{fig3}). Since both transitions lead to the same final
atomic state $|c^{(0)}\rangle _{\text{I}}|c^{(0)}\rangle _{\text{II}}$, one
cannot determine path. A frequency multiplexed entangled single-photon state
will be generated, which has the form of (two frequency components)
\begin{equation}
|1_{\omega _{\text{I}},\omega _{\text{II}}}\rangle =c_{1}|1\rangle _{\omega
_{\text{I}}}|0\rangle _{\omega _{\text{II}}}+c_{2}|0\rangle _{\omega _{\text{%
I}}}|1\rangle _{\omega _{\text{II}}},
\end{equation}%
where $c_{1}=P_{\text{I}}/\sqrt{P_{\text{I}}^{2}+P_{\text{II}}^{2}}$, and $%
c_{2}=P_{\text{II}}/\sqrt{P_{\text{I}}^{2}+P_{\text{II}}^{2}}$. The
maximally entangled single-photon state $|1_{\omega _{\text{I}},\omega _{%
\text{II}}}\rangle =(|1\rangle _{\omega _{\text{I}}}|0\rangle _{\omega _{%
\text{II}}}+|0\rangle _{\omega _{\text{I}}}|1\rangle _{\omega _{\text{II}}})/%
\sqrt{2}$ can be obtained by adjusting the intensities of the write fields
and the mixing ratio of two species atoms. The emitted photon, which is, as
before, only a single one, has imprinted on it the structure of the atomic
excitation: its frequency spectrum consists of the separated lines of finite
width. The scheme also provides an alternative method for measuring level
splittings of different atoms.

Note that in the read process, the atomic excitation is mapped onto a FME
single photon by application of two read fields. The physics behind this
read process is identical to that utilized in electromagnetically induced
transparency (EIT) \cite{Harris97} \textquotedblleft light
storage\textquotedblright\ experiments \cite{Liu}. Due to the suppression of
resonant absorption associated with EIT, the generated entangled
single-photon is not absorbed by the large atomic population in state $%
|s\rangle $. The read laser converts the atomic spin wave into a DSP
\begin{equation}
\hat{\Psi}_{j}(z,t)=\cos \theta _{j}(t)\hat{E}_{S_{j}^{\prime }}-\sin \theta
_{j}(t)\hat{S}_{j},\text{ }(j=\text{I, II}),
\end{equation}%
where $\tan ^{2}\theta _{j}(t)=(g_{j}^{\prime })^{2}N_{j}/\left\vert \Omega
_{R_{j}}\right\vert ^{2}$\ $(j=$I, II$)$, $g_{j}^{\prime }$\ is the
atom-field coupling constants. Under the adiabatic condition, the equation
of motion for $\hat{\Psi}_{j}$ takes a simple form:
\begin{equation}
(\partial _{t}+\upsilon _{g_{j}}\partial _{z})\hat{\Psi}_{j}(z,t)=0.
\end{equation}%
The DSP propagates out of the medium with group velocity $\upsilon _{g}$ and
emerges as a single photon. For a single species three-level atomic system,
the retrievement of a spin wave had been realized in generation of
nonclassical photon pairs \cite{Kuzmich,Wal}. The direction, bandwidth, and
central frequency of the entangled single-photon pulse are determined by the
direction, intensity, and frequency of the retrieve laser \cite{Eisaman}.

Now, we consider the generation efficiency. First, the dark counts of the
single-photon detectors are necessarily considered. The dark count gives a
detector click, but without real photon generation. Commercial actively
quenched single-photon avalanche photodiodes (SPAD) modules that exhibit
dark count rate from 400 Hz to 50 Hz \cite{Commercial}. Extremely low dark
count rate of 5 Hz is reported \cite{Kim}. Second, we must consider the
influence of the finite detection efficiency. In our scheme, according to
four steps operation, the success of generating the frequency multiplexed
entangled single photon $|1_{\omega _{1},\omega _{2}}\rangle $ depends on
the success of generating the desired entangled state $|\phi \rangle $,
which in turn depends on the detection result in step $3$. When no photon is
detected, there is no way of knowing for sure that the entangled state $%
|\phi \rangle $ has been generated or not. Thus the finite detection
efficiency will affect the generation efficiency. When the scheme fails to
generate the desired entangled state, however, the experiment can easily be
repeated for another round of trial within the efficient time of the SPAD.
Conditioned on detecting a single photon, the prepared spin wave is
coherently converted into a FME single photon by applying two read fields $%
\Omega _{R_{\text{1}}}$ and $\Omega _{R_{\text{II}}}$, which is
quasideterministic in ideal case.

In fact, we can improve the generation efficiency by simplifying operation
procedure-the write and read fields are driven on the ensemble according to
the time sequence shown in Fig.~\ref{fig4}. If the single photon generated
by the write field occurs, then the frequency multiplexed entangled single
photon can be generated by the read fields. If not, the read fields are used
as the optical pumping for next round. The frequency multiplexed entangled
single photons can be obtained by repeating the write and read fields
according to the time sequence successively.

As a specific example for realization of our scheme proposed here, we
consider a $^{85}$Rb/$^{87}$Rb mixed atomic ensemble. The energy level
diagrams of the two species atoms is shown in Figs.~\ref{fig2} and \ref{fig3}%
. For the $D_{1}$ line of the Rb-85 species, the$\ \{|s\rangle ,|g\rangle \}$
correspond to the levels $F=\{3,2\}$ of $5S_{1/2}$, and $|e\rangle $
corresponds the level $F=3$ of $5P_{1/2}$, respectively. For the $D_{1}$
line of the Rb-87 species, the$\ \{|s^{\prime }\rangle ,|g^{\prime }\rangle
\}$ correspond to the levels $F=\{2,1\}$ of $5S_{1/2}$, and $|e^{\prime
}\rangle $ corresponds the level $F=1$ of $5P_{1/2}$, respectively. For a mode
match, the write fields $\Omega _{W_{\text{I}}}$ ($\omega _{W_{\text{I}}}$)
and $\Omega _{W_{\text{II}}}$ ($\omega _{W_{\text{II}}}$) are generated
using phase modulation of a single-frequency laser pulse of $\Omega _{W_{3}}$
($\omega _{W_{3}}=(\omega _{es}+\omega _{e^{\prime }s^{\prime }})/2$), where
phase modulation is accomplished by an electro-optical phase modulator,
which produces sidebands with frequencies $\omega _{W_{\text{I}}}=\omega
_{W_{3}}+\delta \omega _{W}$ and $\omega _{W_{\text{II}}}=\omega
_{W_{3}}-\delta \omega _{W}$. Also the the read fields $\Omega _{R_{\text{I}%
}}$ ($\omega _{R_{\text{I}}}$) and $\Omega _{R_{\text{II}}}$ ($\omega _{R_{%
\text{II}}}$) are obtained by another read field of $\Omega _{R_{3}}$ ($%
\omega _{R_{3}}=(\omega _{eg}+\omega _{e^{\prime }g^{\prime }})/2$), which
produces sidebands with frequencies $\omega _{R_{\text{I}}}=\omega
_{R_{3}}-\delta \omega _{R}$ and $\omega _{R_{\text{II}}}=\omega
_{R_{3}}+\delta \omega _{R}$. For the $^{85}$Rb-$^{87}$Rb isotope mixture
system, the relating parameters are shown in Figs. \ref{fig2} and \ref{fig3}%
, then $\Delta =(\omega _{e^{\prime }g^{\prime }}-\omega _{eg})/2=1.368$
GHz, $\delta \omega _{W}=(\omega _{es}-\omega _{e^{\prime }s^{\prime
}})/2=1899.5$ MHz and $\delta \omega _{R}=\Delta =1.368$ GHz.

In conclusion, we present a scheme to realize the entanglement between the
two different species atoms, and then realize the FME single photon based on the entanglement. Injection the write and read fields
according to a certain timing sequence, the generation of FME single photons can be repeated until success is
achieved. Our scheme is feasible under the realistically experimental
conditions.

This work is supported by the National Basic Research Program of China (973
Program) under Grant No.~2011CB921604, the National Natural Science
Foundation of China under Grant No.~10828408, No.~11004058, No.~11004059,
Shanghai Leading Academic Discipline Project under Grant No. B480, and the
Fundamental Research Funds for the Central Universities. \newline
Email:$^{\ast }$zou@iupui.edu; $^{\dag }$wpzhang@phy.ecnu.edu.cn


\begin{thebibliography}{99}
\bibitem{Note} Our scheme also works for chemically different atoms.
\end{thebibliography}
\end{document}